\documentclass[a4paper]{article}
\pdfoutput=1
\usepackage{jheppub}

\usepackage{amsmath}
\usepackage{amssymb}
\usepackage{graphicx,color}

\usepackage{amsmath,amsthm}
\usepackage{txfonts}

\usepackage{bm}

\newcommand{\Ref}[1]{(\ref{#1})}

\newtheorem{Theorem}{Theorem}[section]

\newcommand{\Z}{\mathbb{Z}}
\newcommand{\R}{\mathbb{R}}
\newcommand{\C}{\mathbb{C}}

\newcommand{\half}{\frac{1}{2}}

\newcommand{\Slc}{\mathrm{SL}(2,\mathbb{C})}

\newcommand{\Su}{\mathrm{SU}(2)}

\def\be{\begin{eqnarray}}
\def\ee{\end{eqnarray}}

\newcommand{\cf}{\mathcal F}

\newcommand{\ck}{\mathcal K}

\newcommand{\cm}{\mathcal M}

\newcommand{\cs}{\mathcal S}

\newcommand{\cw}{\mathcal W}

\newcommand{\cz}{\mathcal Z}

\newcommand{\fa}{\mathfrak{a}}  
\newcommand{\fb}{\mathfrak{b}}  
\newcommand{\fc}{\mathfrak{c}}  
\newcommand{\fd}{\mathfrak{d}}

\newcommand{\fs}{\mathfrak{s}}  \newcommand{\Fs}{\mathfrak{S}}

  \newcommand{\Fx}{\mathfrak{X}}
  
\newcommand{\fz}{\mathfrak{z}}

\renewcommand{\a}{\alpha}
\renewcommand{\b}{\beta}
\newcommand{\g}{\gamma}

\newcommand{\eps}{\varepsilon}

\renewcommand{\l}{\lambda}
\renewcommand{\L }{\Lambda}
\renewcommand{\o}{\omega}
\renewcommand{\O}{\Omega}
\renewcommand{\t}{\tau}

\newcommand{\rmd}{\mathrm d}

\newcommand{\lt}{\left}
\newcommand{\rt}{\right}

\newcommand{\lag}{\left\langle}
\newcommand{\rag}{\right\rangle}

\newcommand{\act}{\rhd}
\newcommand{\background}{(\mathring{j}_f,\mathring{g}_{ve},\mathring{z}_{vf})}
\newcommand{\sgn}{\mathrm{sgn}}

\title{On Spinfoam Models in Large Spin Regime}

\author[]{Muxin Han}

\affiliation[]{Centre de Physique Th\'eorique%
\footnote{Unit\'e mixte de recherche (UMR 6207) du CNRS et des Universit\'es
de Provence (Aix-Marseille I), de la Meditarran\'ee (Aix-Marseille II) et du Sud (Toulon-Var); laboratoire affili\'e \`a la FRUMAM (FR 2291).}, CNRS UMR7332, Aix-Marseille Universit\'e and Universit\'e de Toulon, 13288 Marseille, France}

\emailAdd{Muxin.Han(AT)cpt.univ-mrs.fr} 

\abstract{We study the semiclassical behavior of Lorentzian Engle-Pereira-Rovelli-Livine (EPRL) spinfoam model, by taking into account the sum over spins in the large spin regime. We also employ the method of stationary phase analysis with parameters and the so called, almost-analytic machinery, in order to find the asymptotic behavior of the contributions from all possible large spin configurations in the spinfoam model. The spins contributing the sum are written as ${J}_f=\l {j}_f$ where $\l$ is a large parameter resulting in an asymptotic expansion via stationary phase approximation. The analysis shows that at least for the simplicial Lorentzian geometries (as spinfoam critical configurations), they contribute the leading order approximation of spinfoam amplitude only when their deficit angles satisfy $\g\mathring{\Theta}_f\leq\l^{-1/2}$ mod $4\pi\Z$. Our analysis results in a curvature expansion of the semiclassical low energy effective action from the spinfoam model, where the UV modifications of Einstein gravity appear as subleading high-curvature corrections.}


\arxivnumber{}

\begin{document}

\maketitle


\section{Programme of Semiclassical Spinfoam State-Sum}

Loop Quantum Gravity (LQG) is an attempt to make a background independent, non-perturbative
quantization of 4-dimensional General Relativity (GR) -- for reviews, see \cite{book,rev}. Currently the covariant formulation of LQG is understood in terms of Spinfoam Model, which is a sum-over-history of quantum geometries \cite{sfrevs}.  

Here in this paper we study the semiclassical behavior of Lorentzian Engle-Pereira-Rovelli-Livine (EPRL) spinfoam model \cite{EPRL,Carlo} by considering the sum over spins in the large spin regime. The analysis is a natural continuation of the previous studies of large spin asymptotics \cite{semiclassical,CF,HZ,hanPI}, which don't take into account the sum over spins. The result of the present analysis also connects with the recent argument about the ``flatness problem'' proposed in \cite{flatness} where summing over spins is taken into account.   

It is proposed in \cite{hanPI} that the EPRL spinfoam state-sum model \cite{EPRL,Carlo} on a simplical complex $\ck$ has the following path integral representation:
\be
A(\ck)&=&\sum_{J_{f}}d_{J_f} \int_{\Slc}\prod_{\left( v,e\right)}
\rmd g_{ve} \int_{\mathbb{CP}^{1}}\prod_{v\in \partial f} \rmd{z_{vf}}\ e^{S[J_f,g_{ve},z_{vf}]}
\ee
where $f$ labels a triangle in the simplicial complex $\ck$ or a dual face in the dual complex $\ck^*$, $e$ labels a tetrahedron in $\ck$ or an edge in $\ck^*$, and $v$ labels a 4-simplex in $\ck$ or a dual vertex in $\ck^*$.  $J_f$ labels the SU(2) irreps associated to each triangle. $g_{ve}$ is a $\Slc$ group variable associated with each dual half-edge.  $z_{vf}$ is a 2-component spinor. The integrand written into an exponential form $e^S$ with the spinfoam action $S$ written as
\be
S[J_f,g_{ve},z_{vf}]
&=&\sum_{(e,f)}J_f\lt[\ln\frac{\lag Z_{vef},Z_{v'ef}\rag^{2}}{ \lag Z_{v'ef},Z_{v'ef}\rag \lag Z_{vef},Z_{vef}\rag}+i\g \ln \frac{\lag Z_{vef},Z_{vef}\rag}{\lag Z_{v'ef},Z_{v'ef}\rag}\rt]
\label{action}
\ee
where $Z_{vef}=g_{ve}^\dagger z_{vf}$. The spinfoam action $S$ has the following discrete gauge symmetry: Flipping the sign of individual group variable $g_{ve}\mapsto -g_{ve}$ leaves $S$ invariant. Thus the space of group variable is essentially the restricted Lorentz group $\mathrm{SO}^+(1,3)$ rather than its double-cover $\Slc$. $S$ also has the following continuous gauge degree of freedom:
\begin{itemize}
\item Rescaling of each $z_{vf}$\footnote{The measure $\rmd z_{vf}$ is a scaling invariant measure on $\mathbb{CP}^1$.}:
\be
z_{vf}\mapsto \l z_{vf},\ \ \ \l\in\C\setminus\{0\}.
\ee

\item $\Slc$ gauge transformation at each vertex $v$:
\be
g_{ve}\mapsto x^{-1}_vg_{ve},\ \ \ z_{vf}\mapsto x^\dagger_vz_{vf},\ \ \ x_v\in\Slc.
\ee

\item SU(2) gauge transformation on each edge $e$:
\be
g_{ve}\mapsto g_{ve}h_e^{-1},\ \ \ h_e\in\Su.
\ee
\end{itemize}

We can make the following parametrization of the 2-spinor $z_{vf}^\a$ and the group variables $g_{ve}\in\Slc$ in the spinfoam action $S$:
\be
z_{ve}^\a=\left(\begin{array}{c}1\\ \fz_{vf} \end{array}\right)\ \ \text{and}\ \ \
g_{ve}=\left(\begin{array}{cc}\fa_{ve} & \fc_{ve}\\ \fb_{ve} & \fd_{ve} \end{array}\right),\ \ \ \ \fa_{ve}\fd_{ve}-\fc_{ve}\fb_{ve}=1
\ee
where $\fz_{ve},\fa_{ve},\fb_{ve},\fc_{ve},\fd_{ve}\in\C$. The arguments of the logarithmics in Eq.\Ref{action} are two ratios of the polynomials in the variables $\fz_{ve},\fa_{ve},\fb_{ve},\fc_{ve},\fd_{ve}$ and their complex conjugates. If we write the $\Slc$ and $\Su$ gauge transformations by 
\be
x^\dagger_v=\left(\begin{array}{cc}\a_{v} & \b_{v}\\ \g_{v} & \delta_{v} \end{array}\right),\ \ \  h_e=\left(\begin{array}{cc}\mu_{e} & -\bar{\nu}_{e}\\ \nu_{e} & \bar{\mu}_{e} \end{array}\right)
\ee
then $\fz_{ve},\fa_{ve},\fb_{ve},\fc_{ve},\fd_{ve}$ transform in the following way:
\be
\fz_{ve}\mapsto\frac{\g_v+\delta_{v} \fz_{vf}}{\a_v+\b_v \fz_{vf}},\ \ \text{and}\ \ 
\left(\begin{array}{cc}\fa_{ve} & \fc_{ve}\\ \fb_{ve} & \fd_{ve} \end{array}\right)\mapsto\left(\begin{array}{cc}\bar{\delta}_{v} & -\bar{\g}_{v}\\ -\bar{\b}_{v} & \bar{\a}_{v} \end{array}\right)\left(\begin{array}{cc}\fa_{ve} & \fc_{ve}\\ \fb_{ve} & \fd_{ve} \end{array}\right)\left(\begin{array}{cc}\bar{\mu}_{e} & \bar{\nu}_{e} \\ -{\nu}_{e} & {\mu}_{e} \end{array}\right).
\ee

For the convenience of the discussion, we define the notion of the partial-amplitude $A_{j_f}(\ck)$ by collecting all the integrations
\be
A_{J_f}(\ck):=\int
\rmd g_{ve} \int\rmd z_{vf}\ e^{S\lt[J_f,g_{ve},z_{vf}\rt]}  \label{Aj}
\ee
So that the spinfoam state-sum model is given by a sum of partial amplitude over all the spin configurations $\{J_f\}_f$ on the simplicial complex $\ck$ 
\be
A(\ck)=\sum_{J_f}d_{J_{f}}A_{J_f}(\ck).
\ee
Note that the infinite spin-sum in $A(\ck)$ may result in a divergent result. A way to regularizing the spin-sum is to replace $\Slc$ in the definition by the quantum group $\mathrm{SL}_q(2,\C)$ \cite{QSF}, which also relates to the cosmological constant term in spinfoam formulation \cite{QSFasymptotics}.

So far the semiclassical properties of the spinfoam model are mostly understood via the asymptotic analysis of the partial amplitude $A_J(\ck)$ as all the spins $J_f=\l j_f$ are uniformly large, where $\l$ is a large parameter to scale the spins uniformly and $j_f\sim o(1)$. In such a context, the analysis in \cite{hanPI,HZ}\footnote{See also \cite{CF} concerning Euclidean spinfoam models.} shows that given a Regge-like\footnote{A Regge-like spin configuration $\{j_f\}_f$ means that $\g j_f$ for each $f$ can be interpreted as the area of the triangle dual to $f$, i.e. there exists a set of edge-lengths $\{s_\ell\}_\ell$ for the edges of the simplicial complex, such that $\{\g j_f\}_f$ and $\{s_\ell\}_\ell$ satisfy the relation between areas and edge-lengths for all triangles.} spin configuration $\{j_f\}_f$ on the simplicial complex $\ck$. A critical configuration $(j_f,g_{ve},z_{vf})$\footnote{$j_f$ is parameter in the analysis of partial amplitude $A_J(\ck)$ since it hasn't been summed yet.} satisfying $\Re(S)=\delta_{g}S=\delta_z S=0$ and a nondegeneracy condition\footnote{The nondegneracy condition is an assumption for the 5 group variables $g_{ve}$ at each $v$:
\be
\prod_{e_1,e_2,e_3,e_4=1}^5\det\Big(N_{e_1}(v),N_{e_2}(v),N_{e_3}(v),N_{e_4}(v)\Big)\neq0\ \ \ \ \text{where}\ \ \ N_e(v)= g_{ve}\act(1,0,0,0)^t.\label{proddet}
\ee} has a canonical geometric interpretation as a classical simplicial Lorentzian geometry on $\ck$. More precisely, there is an equivalence between spinfoam critical data and the geometrical data (see \cite{hanPI} for details):
\be
\text{Critical configuration}\ (j_f,g_{ve},z_{vf})\ \text{with nondegenercy} \Longleftrightarrow (\pm_vE_\ell(v),\eps)\label{equivalence}
\ee 
$E_\ell(v)$ is a set of edge-vectors, or namely, a discrete cotetrad, assigned to the simplical complex $\ck$, up to global sign ambiguity $\pm_v$ at each $v$. $E_\ell(v)$ specifies the Lorentzian simplicial geometry on $\ck$, and results in nonvanishing 4-volume $V_4(v)$ of all geometrical 4-simplices. $\eps=\pm1$ is a global sign ambiguity on the simplicial complex, which can be fixed by the boundary data if there is a boundary of the spinfoam. 

The set of critical configurations contributes the leading asymptotic behavior of the partial amplitude $A_{J_f}(\ck)$ as the spins $J_f=\l j_f$ become large uniformly for all $f$. If we consider the (physical) critical configurations which corresponds geometrically to the globally oriented and time-oriented spacetimes\footnote{The notion of oriented and time-oriented spacetimes for spinfoam critical data is defined in \cite{hanPI} by the equivalence Eq.\ref{equivalence}. A discrete oriented spacetime requires the oriented 4-volume $V_4(v)$ from cotetrad is uniformly positive/negative. A time-oriented spacetime requires the discrete spin connection along a loop belongs to $\mathrm{SO}^+(1,3)$. These notions are independent of the $\pm_v$ ambiguity.}, these critical configurations give the Regge action of discrete GR at the leading order of the $1/\l$ asymptotic expansion.

In the present analysis, we study the spinfoam state-sum $A(\ck)$ perturbatively by using the background field method, in the regime that $J_f=\l j_f$ is uniformly large, where $\l$ is a fixed large parameter (smaller than the cut-off $\L$). The perturbative analysis is performed in a neighborhood of a given critical configuration $\background$ of $S$ as the background/vacuum, which corresponds to the globally oriented and time-oriented spacetime. The background spins are $\mathring{J}_f=\l \mathring{j}_f$ with $\mathring{j}_f\sim o(1)$. Sometime we also refer $\l$ to be the (averaged) background spin. Given the critical configuration $\background$ as the vacuum, we consider the fluctuations of all the variables in a neighborhood\footnote{The full nonperturbative spinfoam model can be understood as a sum of the perturbation theories over different vacuums. } at $\background$ in the space $\cm_{j,g,z}$ of spinfoam variables $(j_f,g_{ve},z_{vf})$\footnote{We shall see that we have to consider the complexified space $\cm_{j,g,z}^\C$ of the spinfoam variables in order to study the sum over $j_f$.}. 

In our context $j_f$ is a perturbation of the background $\mathring{j}_f$. The gap of two neighboring spins is $\Delta j_f=\frac{1}{2\l}$ while $\Delta J_f=\half$. The spinfoam action $S[J_f,g_{ve},z_{vf}]$ is written as $\l S[j_f,g_{ve},z_{vf}]$ since $S$ is linear to $J_f$. The formula of $S_f$ is identical to Eq.\Ref{action} with $J_f$ replaced by $j_f$. 

In order to analyze the semiclassical behavior of the spinfoam state-sum, we recall the spinfoam state-sum written in the following form 
\be
A(\ck)=\sum_{j}d_{\l j}\int\rmd\mu(x)\ e^{\l S[j,x]},\ \ \ \text{with}\ \ \ x=(g_{ve},z_{vf})\label{PI}
\ee
The semiclassical behavior of the above state-sum is analyzed in the following by assuming the background spin $\l$ is a fixed large but finite parameter, and expand the summand $\int\rmd\mu\ e^{\l S}$ perturbatively into an asymptotic series of $1/\l$ for all possible spin configurations. Such an expansion gives us an effective action $\cw_\ck[J_f]$ for the spin variables as an asymptotic series. Then the sum of the $\exp\cw_\ck[J_f]$ over all spins gives us the information about semiclassical behavior of the state-sum. 




It is clear that the $1/\l$ asymptotic analysis of the partial amplitude $A_{\l j_f}(\ck)$ depends on the spin parameters $j_f$, which are not integrated in $A_{\l j_f}(\ck)$. The parameter $j_f$ has control about whether the critical equations $\Re(S)=\delta_gS=\delta_zS=0$ has solutions or not. Given a neighborhood $\O$\footnote{$\O$ is a compact neighborhood in the space of spinfoam configurations $\cm_{j,g,z}=\R_+^{\#_f}\times \Slc^{\#_{(e,v)}}\times (\mathbb{CP}^1)^{\#_{(v,f)}}$ modulo gauge transformations.} at $\background$. Some values of $j_f$'s doesn't result in solution of the critical equations $S'=0$ and $\Re(S)=0$ \cite{hanPI,HZ,CF}. 
Here we call these values of $j_f$ \emph{Non-Regge-like}, otherwise we call them \emph{Regge-like}. A Regge-like spin configuration $j_f$ admits the critical configuration $(j_f,g_{ve},z_{vf})$ which can be interpreted as nondegenerate geometry, assuming this neighborhood $\O$ doesn't tough any critical configuration of degenerate geometry\footnote{The degeneracy restricts of the group variables $g_{ve}$ into a lower-dimensional submanifold. Thus such a neighborhood always exists.}. The Non-Regge-like spin configurations in $\O$ doesn't result in any critical point. 

According to the general result of generalized stationary phase approximation (Theorem 7.7.5 and 7.7.1 in \cite{stationaryphase}), In case there is no critical point in the region of integral $\int_K e^{\l S}\rmd\mu$, 
\be
\lt|\int_K e^{\l S(x)}\rmd \mu(x)\rt|\leq C\lt(\frac{1}{\l}\rt)^k\sup_K\frac{1}{\lt(|S'|^2+\Re(S)\rt)^{k}}
\ee
the integral decays faster than $(1/\l)^k$ for all $k\in\Z_+$, provided that $\sup([|S'|^2+\Re(S)]^{-k})$ is finite (i.e. doesn't cancel the $(1/\l)^k$ behavior in front).

For the Non-Regge-like spins $j_f$, however, it is \emph{not} necessary that the corresponding $A_{\l j_f}(\ck)$ decays faster than $(1/\l)^k$ for all $k\in\Z_+$. As an example, for a non-Regge-like $j_f$ very close to a Regge-like $j_f$ with the minimal gap $\Delta j_f=\frac{1}{2\l}$, $\sup([|S'|^2+\Re(S)]^{-k})$ is likely to be large and cancel the $(1/\l)^k$ behavior (such a cancellation is described more precisely in Section \ref{EFF}). Therefore if we consider the semiclassical behavior of the state-sum $A(\ck)=\sum_{J_f}d_{J_f}A_{J_f}(\ck)$ in the large spin regime, we should in general not ignore these ``almost Regge-like'' contributions. A general way is presented in Sections \ref{analytic} and \ref{EFF} to obtain an $1/\l$ asymptotic expansion, taking into account the contributions from all the spins, both Regg-like and non-Regge-like.

\section{Large Spin Effective Theory}

Given that the spinfoam state-sum is a finite sum of the partial amplitude:
\be
A(\ck)=\sum_{J_f} d_{J_f}A_{J_f}(\ck)
\ee
If we define an ``Spin Effection Action'' $\cw_\ck[J_f]$ by
\be
A_{J_f}(\ck)=\exp\cw_\ck[J_f]
\ee
Then the formally the spinfoam state-sum is effectively a partition function of a spin system
\be
A(\ck)=\sum_{J_f} d_{J_f}\exp\cw_\ck[J_f]. 
\ee
Here we study the spin effective action $\cw_\ck[J_f]$ as an asymptotic expansion in terms of $1/\l$. Thus the effective theory of spinfoam state-sum is studied in the large spin regime. 

In the large spin regime, by the integral expression of the partial amplitude, the spin effective action can be written as 
\be
\cw_\ck[\l j_f]=\ln\int
\rmd g_{ve} \int\rmd z_{vf}\ e^{\l S\lt[j_f,g_{ve},z_{vf}\rt]} 
\ee
We would like to compute $\cw_\ck[\l j_f]$ perturbatively at the background spin configuration $\mathring{J}_f=\l\mathring{j}_f$. So $\cw_\ck[\l j_f]$ can be written as a power series of the perturbations $\fs_f:=j_f-\mathring{j}_f$:
\be
\cw_\ck[\l j_f]=\l\lt[\cw_0+\sum_f\cw_1^f\fs_f+\sum_{f,f'}\cw_2^{f,f'}\fs_{f}\fs_{f'}+\sum_{f,f',f''}\cw_3^{f,f',f''}\fs_{f}\fs_{f'}\fs_{f''}+o(\fs_f^4)\rt]\label{SEA}
\ee
where the coefficients is given by
\be
\l\cw_{n}^{f_1,\cdots,f_n}=\frac{\partial^n}{\partial j_{f_1}\cdots\partial j_{f_n}}\lt[\ln\int
\rmd g_{ve} \int\rmd z_{vf}\ e^{\l S\lt[j_f,g_{ve},z_{vf}\rt]}\rt]_{j_f=\mathring{j}_f}.
\ee

Each of the coefficients $\cw_n$ can be computed as an $1/\l$ asymptotic series, by using the generalized stationary phase analysis guided by the following general result (Theorem 7.7.5 in \cite{stationaryphase}):

\begin{Theorem}\label{asymptotics}
Let $K$ be a compact subset in $\R^N$, $X$ an open neighborhood of $K$, and $k$ a positive integer. If (1) the complex functions $u\in C^{2k}_0(K)$, $S\in C^{3k+1}(X)$ and $\Re(S)\leq 0$ in $X$; (2) there is a unique point $x_0\in K$ satisfying $\Re(S)(x_0)=0$, $S'(x_0)=0$, and $\det S''(x_0)\neq 0$. $S'\neq0$ in $K\setminus \{x_0\}$, then we have the following estimation:
\be
\lt|\int_K u(x)e^{\l S(x)}\rmd x-e^{\l S(x_0)}\lt(\frac{2\pi}{\l}\rt)^{\frac{N}{2}}\frac{e^{\mathrm{Ind}(S'')(x_0)}}{\sqrt{\det(S'')(x_0)}}\sum_{s=0}^{k-1}\lt(\frac{1}{\l}\rt)^s L_s u(x_0)\rt|\leq C\lt(\frac{1}{\l}\rt)^{k+\frac{n}{2}}\sum_{|\a|\leq 2k}\sup\lt|D^\a u\rt|
\ee
Here the constant $C$ is bounded when $f$ stays in a bounded set in $C^{3k+1}(X)$. We have used the standard multi-index notation $\a =\lag\a_1,\cdots,\a_n\rag$ and
\be
D^\a=\frac{\partial^{|\a|}}{\partial x_1^{\a_1}\cdots\partial x_n^{\a_n}},\ \ \ \ \text{where}\ \ \ \ |\a|=\sum_{i=1}^n\a_i
\ee
$L_s u(x_0)$ denotes the following operation on $u$:
\be
L_s u(x_0)=i^{-s}\sum_{l-m=s}\sum_{2l\geq 3m}\frac{2^{-l}}{l!m!}\lt[\sum_{a,b=1}^NH^{-1}_{ab}(x_0)\frac{\partial^2}{\partial x_a\partial x_b}\rt]^l\lt(g_{x_0}^m u\rt)(x_0)\label{Lsu}
\ee
where $H(x)=S''(x)$ denotes the Hessian matrix and the function $g_{x_0}(x)$ is given by
\be
g_{x_0}(x)=S(x)-S(x_0)-\frac{1}{2}H^{ab}(x_0)(x-x_0)_a(x-x_0)_b
\ee
such that $g_{x_0}(x_0)=g_{x_0}'(x_0)=g_{x_0}''(x_0)=0$.

\end{Theorem}
\noindent
Note that in Eq.\Ref{Lsu}, the expression of $L_s u(x_0)$ only sums a finite number of terms for all $s$. For each $s$, $L_s$ is a differential operator of order $2s$ acting on $u(x)$. For example we list the possible types of terms in the sums corresponding to $s=1$ and $s=2$
\begin{itemize}

\item In the case $s=1$, the possible $(m,l)$ are $(m,l)=(0,1),(1,2),(2,3)$ to satisfy $2l\geq 3m$. The corresponding terms are of the types
\be
(m,l)=(0,1):&& \partial^2u(x_0)\nonumber\\
(m,l)=(1,2):&& \partial^3g_{x_0}(x_0)\partial u(x_0),\ \ \partial^4g_{x_0}(x_0)u(x_0) \nonumber\\
(m,l)=(2,3):&&\partial^3g_{x_0}(x_0)\partial^3g_{x_0}(x_0) u(x_0)
\ee
where the indices of $\partial$ are contracted with the Hessian matrix $H(x_0)$.

\item In the case $s=2$, the possible $(m,l)$ are $(m,l)=(0,2),(1,3),(2,4),(3,5),(4,6)$ to satisfy $2l\geq 3m$. The corresponding terms are of the types
\be
(m,l)=(0,2):&& \partial^4u(x_0)\nonumber\\
(m,l)=(1,3):&& \partial^pg_{x_0}(x_0)\partial^q u(x_0),\ \ \ \
(p\geq 3,\ p+q=6) \nonumber\\
(m,l)=(2,4):&& \partial^{p_1}g_{x_0}(x_0)\partial^{p_2}g_{x_0}(x_0) \partial^qu(x_0)\ \ \ \
(p_1,p_2\geq 3.\ p_1+p_2+q=8)\nonumber\\
(m,l)=(3,5):&& \partial^{p_1}g_{x_0}(x_0)\partial^{p_2}g_{x_0}(x_0)\partial^{p_3}g_{x_0}(x_0) \partial^qu(x_0)\ \ \ \  (p_1,p_2,p_3\geq 3.\ p_1+p_2+p_3+q=10)\nonumber\\
(m,l)=(4,6):&& \partial^{3}g_{x_0}(x_0)\partial^{3}g_{x_0}(x_0)\partial^{3}g_{x_0}(x_0)\partial^{3}g_{x_0}(x_0) u(x_0)
\ee
where the indices of $\partial$ are contracted with the Hessian matrix $H(x_0)$.

\end{itemize}

The spin effective action $\cw_\ck$ is evaluated at a compact neighborhood of $K$ a critical $(g_{ve},z_{vf})$, i.e. we impose a smooth function $u(g_{ve},z_{vf})$ of compact support $K$ such that
\be
e^{\cw_\ck[\l j_f]}u\lt(\mathring{g}_{ve},\mathring{z}_{vf}\rt)=\int\rmd g_{ve}\rmd z_{vf}\,e^{\l S[j_f,g_{ve},z_{vf}]}u\lt(g_{ve},z_{vf}\rt)
\ee  
The compact support function $u(g,z)$ comes from a partition of unity: i.e. a collection of smooth function $0\leq u_\mu(g,z)\leq 1$ of compact support, such that (1) $\sum_\mu u_{\mu}(g,z)=1$ (2) the support $K_\mu$ of each of them containing at most one critical configuration\footnote{In the present paper we only consider the sector of geometric critical configurations, which clearly contains a finite number of isolated critical points (modulo the gauge obits) \cite{hanPI}. }. So
\be
A_{J_f}(\ck)=\sum_\mu\int\rmd g_{ve}\rmd z_{vf}\,e^{\l S[j_f,g_{ve},z_{vf}]}u_\mu\lt(g_{ve},z_{vf}\rt).
\ee

Employing Theorem \ref{asymptotics}, we can evaluate each coefficient $\cw_n$ in Eq.\Ref{SEA} perturbatively at the large spin regime. Here we denote the background data by $\mathring{x}=\background\Leftrightarrow(\pm\mathring{E}_\ell(v),\mathring{\eps}=-1)$ (without loss of generality, we set the global sign $\eps=-1$ for convenience).
\begin{description}
\item[\underline{0th Order}:] If we denote by $N_{g,z}$ the number of degrees of freedom for $g_{ve},z_{vf}$ modulo gauge, and denote by $\mu(x)$ the Jacobian between $\rmd g_{ve}\rmd z_{vf}$ and the Lebesgue measure $\rmd^N x$, then we have
\be
\l \cw_0=\l S[\mathring{x}]-\frac{N_{g,z}}{2}\ln{\l}+\ln\lt[(2\pi)^{\frac{N_{g,z}}{2}}\frac{e^{\mathrm{Ind}(S'')(\mathring{x})}}{\sqrt{\det(S'')(\mathring{x})}}\mu(\mathring{x})\rt]+o(\l^{-1})\label{cw0}
\ee 
where the leading order contribution $S[\mathring{x}]$ is the Lorentzian Regge action on the simplicial complex $\ck$ with the triangle area given by $\g \mathring{j}_f$:
\be
S[\mathring{x}]=i\sum_f\g \mathring{j}_f\mathring{\Theta}_f
\ee
which is resulting from the asymptotic analysis in \cite{hanPI,HZ}. $\mathring{\Theta}_f$ denotes the deficit/dihedral angle in the bulk/boundary at the triangle $f$ from the discrete background geometry $\mathring{E}_\ell(v)$.

\item[\underline{1st Order}:] If we write the action $S[j_f,g_{ve},z_{vf}]=\sum_f j_f\cf_f[g_{ve},z_{vf}]$, then
\be
\l \cw_1^f&=&\frac{\partial}{\partial j_f}\ln\int
\rmd g_{ve} \int\rmd z_{vf}\ e^{\l S\lt[j_f,g_{ve},z_{vf}\rt]}=\l\frac{\int
\rmd g_{ve} \int\rmd z_{vf}\ \cf_f[g_{ve},z_{vf}]\ e^{\l S\lt[j_f,g_{ve},z_{vf}\rt]}}{\int
\rmd g_{ve} \int\rmd z_{vf}\ e^{\l S\lt[j_f,g_{ve},z_{vf}\rt]}}\nonumber\\
&=&i\l\g\mathring{\Theta}_f-i\g\mathring{\Theta}_f\frac{L_1 (u\mu)(\mathring{x})}{u\mu(\mathring{x})}+\frac{L_1 \lt(u\mu\cf_f\rt)(\mathring{x})}{u\mu(\mathring{x})}+o(\l^{-1})\label{cw1}
\ee
where the critical value $\cf_f[\mathring{g}_{ve},\mathring{z}_{vf}]$ is given by $i\g\mathring{\Theta}_f$.

\item[\underline{2nd Order}:] $\l \cw_2^{f,f'}$ can be expressed by a connected correlation function of $\cf_f[g_{ve},z_{vf}]$ 
\be
\l \cw_2^{f,f'}
&=&\l^2\frac{\int \cf_f\cf_{f'}\ e^{\l S\lt[j_f,g_{ve},z_{vf}\rt]}}{\int
 e^{\l S\lt[j_f,g_{ve},z_{vf}\rt]}}-\l^2
 \frac{\int
 \cf_f\ e^{\l S\lt[j_f,g_{ve},z_{vf}\rt]}}{\int
 e^{\l S\lt[j_f,g_{ve},z_{vf}\rt]}}\frac{\int \cf_{f'}\ e^{\l S\lt[j_f,g_{ve},z_{vf}\rt]}}{\int
 e^{\l S\lt[j_f,g_{ve},z_{vf}\rt]}}\nonumber\\
&=&\l \lt[\half\sum_{a,b=1}^{N_{g,z}}H^{-1}_{ab}\partial_a\cf_f\partial_b\cf_{f'}\rt](\mathring{x})+o(1).\label{cw2}
\ee
where $H$ is the Hessian matrix of the action $S$. The computation of $\cw_2^{f,f'}$ is similar to the formalism of graviton propagator computation in LQG, see e.g.\cite{propagator}. Notice that $\partial_{z_{vf}}\cf_f(\mathring{x})=0$ which is nothing but the equation of motion satisfied by the critical configurations. Thus
\be
\sum_{a,b=1}^{N_{g,z}}H^{-1}_{ab}\partial_a\cf_f\partial_b\cf_{f'}(\mathring{x})=\sum_{g,g'}H^{-1}_{g,g'}\partial_g\cf_f\partial_{g'}\cf_{f'}(\mathring{x})
\ee
where $g,g'$ denots the group variables $g_{ve},g_{v'e'}$. We compute $\cw_2^{f,f'}$ by using the expression of Hessian matrix in \cite{HZ,propagator}. As a result, $\cw_2^{f,f'}$ is local in the sense that it vanishes unless the triangles $f,f'$ belong to the same tetrahedron $e$. The nonvanishing $\cw_2^{f,f'}$ reads
\be
\cw_2^{f,f'}=\frac{2(1+2i\g-4\g^2-2i\g^3)}{5+2i\g}\hat{n}_{ef}^t\mathbf{X}^{-1}_e\hat{n}_{ef'}\label{CW2}
\ee
where the matrix $\mathbf{X}_e$ is given by
\be
\mathbf{X}_e^{ij}\equiv\sum_fj_f\lt(-\delta^{ij}+\hat{n}^i_{ef}\hat{n}_{ef}^j+i\eps^{ijk}\hat{n}_{ef}^k\rt)
\ee
Here the unit 3-vector $\hat{n}_{ef}$ determined by $\background$ is the normal vector of the triange $f$ in the frame of the tetrahedron $e$ (see \cite{hanPI}).

\item[\underline{3rd Order}] We write $\l \cw_3^{f_1,f_2,f_3}$ in terms of a connected correlation function $\lag\cf_{f_1}\cf_{f_2}\cf_{f_3}\rag_{\text{connect}}$:
\be
\l \cw_3^{f_1,f_2,f_3}&=&\l^3\Big[\lag\cf_{f_1}\cf_{f_2}\cf_{f_3}\rag-\lag\cf_{f_1}\cf_{f_2}\rag\lag\cf_{f_3}\rag-\lag\cf_{f_1}\cf_{f_3}\rag\lag\cf_{f_2}\rag-\lag\cf_{f_2}\cf_{f_3}\rag\lag\cf_{f_1}\rag\nonumber\\
&&+2\lag\cf_{f_1}\rag\lag\cf_{f_2}\rag\lag\cf_{f_3}\rag\Big]
\ee
where e.g. $\lag\cf_{f_1}\cf_{f_2}\cf_{f_3}\rag={\int\cf_{f_1}\cf_{f_2}\cf_{f_3}\ e^{\l S\lt[j_f,g_{ve},z_{vf}\rt]}}\big/{\int
 e^{\l S\lt[j_f,g_{ve},z_{vf}\rt]}}$. It is shown in general in the LQG 3-point function computation \cite{3pt} that the leading contribution of $\lag\cf_{f_1}\cf_{f_2}\cf_{f_3}\rag_{\text{connect}}$ is of order $o(\l^{-2})$. Thus the leading contribution of $\l \cw_3^{f_1,f_2,f_3}$ is again of order $\l$, i.e.
\be
\l \cw_3^{f_1,f_2,f_3}&=&\l\lt[-g^{abc}\cf^i_{f_1}\cf^j_{f_2}\cf^k_{f_3}H^{-1}_{ai}H^{-1}_{bj}H^{-1}_{ck}+\lt(\cf^{ij}_{f_1}\cf^k_{f_2}\cf^l_{f_3}+\cf^k_{f_1}\cf^{ij}_{f_2}\cf^l_{f_3}+\cf^k_{f_1}\cf^l_{f_2}\cf^{ij}_{f_3}\rt)H^{-1}_{ik}H^{-1}_{jl}\rt](\mathring{x})\nonumber\\
&&+o(1) 
\ee
where $g^{abc}\equiv \partial_a\partial_b\partial_c g_{\mathring{x}}$, $\cf_f^a\equiv\partial_a\cf_f$, and $\cf_f^{ab}\equiv\partial_a\partial_b\cf_f$. 

\end{description}

The above computation suggests that the coefficients $\cw_n$ may be written as an asymptotic series in terms of $\l^{-1}$:
\be
\cw_n=W_n^{(0)}+\lt(\frac{1}{\l}\rt)W_n^{(1)}+\lt(\frac{1}{\l}\rt)^2W_n^{(2)}+o(\l^{-3})\label{cwn}
\ee
where $W_n^{(k)}\sim o(1)$. Such an expansion is illustrated by the above computation up to $\cw_3$, and will be shown to arbitrary $\cw_n$ ($n>0$) by the application of almost-analytic machinery in Section \ref{EFF}. The only exception $\cw_0$ contains a term proportional to $\ln\l$ in its asymptotic expansion. Thus we write $W_0^{(0)}=i\sum_f\g \mathring{j}_f\mathring{\Theta}_f-\frac{N_{g,z}}{2\l}\ln\l$ so that Eq.\Ref{cwn} is valid for all $n$.

We define the perturbative spinfoam state-sum in the large spin regime with the spin effection action given by Eq.\Ref{SEA} ($N_f$ denotes the number of triangles in $\ck$)
\be
A_\t(\ck)
&=&(2\l)^{N_f}e^{\l\cw_0}\sum_{\fs_f=-\infty}^\infty\prod_{f}\lt(\fs_f+\mathring{j}_f+\frac{1}{2\l}\rt) e^{\l\sum_f\cw_1^f\fs_f+\l\sum_{f,f'}\cw_2^{f,f'}\fs_f\fs_{f'}+o(\fs^3)}\t\lt(\fs_f\rt).
\ee
where $\t\lt(\fs_f\rt)$ is a compact support smooth function on $\R^{N_f}$ which comes from a partition of unity \cite{stationaryphase} $\sum_I\t_I(j_f)=1$ on the space of $j_f$. So the spinfoam state-sum in the large spin regime can be written as
\be
A(\ck)=\sum_I\sum_{j_f=-\infty}^\infty d_{J_f}\t_I(j_f)\exp\cw_\ck[\l j_f].
\ee
If we define the spin fluctuation $\Fs_f=\l\fs_f=J_f-\mathring{J}_f$ with $\Delta\Fs_f=\half$, then
\be
A_\t(\ck)&=&e^{\l\cw_0}\sum_{\Fs_f=-\infty}^\infty\prod_{f}\lt(2\Fs_f+2\mathring{J}_f+1\rt) e^{\sum_f\cw_1^f\Fs_f+\frac{1}{\l}\sum_{f,f'}\cw_2^{f,f'}\Fs_f\Fs_{f'}+o(\l^{-3}\Fs^3)}\t\lt(\Fs_f/\l\rt).\label{spinsystem}
\ee
Therefore as an effective theory, the spinfoam state-sum may be viewed analogously as a partition function of a (classical) spin system with the (complex) Hamiltonian 
\be
\sum_f\cw_1^f\Fs_f+\frac{1}{\l}\sum_{f,f'}\cw_2^{f,f'}\Fs_f\Fs_{f'}+\frac{1}{\l^2}\sum_{f,f'}\cw_3^{f,f',f''}\Fs_f\Fs_{f'}\Fs_{f''}+o(\l^{-4}\Fs_f^4)
\ee
where $\cw_1^f=i\g\mathring{\Theta}_f+o(\l^{-1})$ is an external (complex) source in the system, and the complex coefficients $\cw_n^{f_1\cdots f_n}$ defines the interactions between the spin fluctuations $\Fs_f$ located at each triangle $f$ of the simplicial complex $\ck$.

Although the quadratic term $\cw_2^{f,f'}\Fs_f\Fs_{f'}$ has been shown to be local ($\cw_2^{f,f'}$ vanishes unless the trangles $f,f'$ belong to the same tetrahedron $e$), the interactions $\cw_n^{f_1\cdots f_n}$ is in general nonlocal, i.e. the spin $\Fs_f$ localed at the triangle $f$ has the interactions not only with neighboring spins, but also with many other spins located at the triangles far away from $f$. It is because the Hessian matrix $H$ with respect to the variables $g_{ve},z_{vf}$ is not block-diagonal (see \cite{HZ} for a computation of Hessian matrix), the inverse $H^{-1}$ may have nonzero entries for all the matrix elements\footnote{In the spinfoam action $S$, the $(j_f, g_{ve}, z_{vf})$-variables located far away from each other on $\ck$ doesn't correlate. The nonzero matrix elements in the Hessian matrix $H$ are close to diagonal, although $H$ is not block-diagonal. As an example, we consider a tridiagonal matrix with its inverse
\be
T=\begin{pmatrix}
a_1\ & b_1 \\
c_1\ & a_2 &\ b_2 \\
&\ c_2 & \ddots & \ddots \\
& & \ddots & \ddots &\ b_{n-1} \\
& & &\ c_{n-1} &\ a_n
\end{pmatrix}\ \ \text{and}\ \  
(T^{-1})_{ij} = \begin{cases}
(-1)^{i+j}b_i \cdots b_{j-1} \theta_{i-1} \phi_{j+1}/\theta_n &\ \text{ if } i \leq j\\
(-1)^{i+j}c_j \cdots c_{i-1} \theta_{j-1} \phi_{i+1}/\theta_n &\ \text{ if } i > j\\
\end{cases}
\ee
where $\{\theta_i\}_i$ satisfy the recurrence relation $\theta_i = a_i \theta_{i-1} - b_{i-1}c_{i-1}\theta_{i-2} \text{ for } i=2,3,\ldots,n$ with initial conditions $\theta_0 = 1$, $\theta_1 = a_1$. $\{\phi_i\}_i$ satisfy $\phi_i = a_i \phi_{i+1} - b_i c_i \phi_{i+2} \text{ for } i=n-1,\ldots,1$ with the initial conditions $\phi_{n+1} = 1$ and $\phi_n = a_n$ \cite{tridiagonal}.
}.

\section{(Almost)-analytic Machinery}\label{analytic}

In order to understand the semiclassical behavior of the spinfoam state-sum, we should perform the spin sum for $A(\ck)$ in the spin partition function Eq.\Ref{spinsystem}. It relies on a more detailed understanding of the functional properties of spin effective action $\cw_\ck[J_f]$. In particular, Eq.\Ref{cwn} should be shown for general $n$.

Let's come back to the asymptotic analysis of the partial amplitude $A_{J_f}(\ck)=\exp\cw_{\ck}[J_f]$ within a neighborhood of the large background spin $\mathring{J}_f=\l \mathring{j}_f$. The partial amplitude $A_{\l j_f}(\ck)$ Eq.\Ref{Aj} is of the form\footnote{The integral is convergent once we gauge-fix the $\Slc$ gauge freedom at each vertex, see \cite{EP} for details.}
\be
\int e^{\l S[j_f,g_{ve},z_{vf}]}\rmd\mu[g_{ve},z_{vf}]\label{Aj1}
\ee
where we see that the spins $j_f$ enters as free parameters and the asymptotics of $A_{\l j_f}(\ck)$ clearly depends on $j_f$. The analysis in \cite{hanPI,HZ,CF} applies the generalized stationary phase approximation (Theorem \ref{asymptotics}) to the above type of integral (the method generalizes the standard stationary phase approximation to the case of complex action). It is shown that the leading order asymptotics of the integral Eq.\Ref{Aj1} are contributed by the \emph{critical configurations} $(j_f,g_{ve},z_{vf})$ which solves the critical equations $\delta_{g,z}S=0$ and $\Re (S)=0$. However there is a difficulty in the generalized stationary phase analysis of the complex action $S[j_f,g_{ve},z_{vf}]$ with $j_f$ as parameter. Not all values of $j_f$ results in solutions of the critical equations $\delta_{g,z}S=0$ and $\Re (S)=0$ in a neighborhood of $\background$. The values of $j_f$ which result in solutions are Regge-like spin configurations, which can be interpreted geometrically as triangle areas. For the non-Regge-like spins $j_f$ which don't result in critical solutions, the integral Eq.\Ref{Aj1} decays as $\l$ becomes large in the following way:
\be
\lt|A_{\l j_f}(\ck)\rt|\leq \lt(\frac{1}{\l}\rt)^N\frac{C}{\min\lt(|\delta_{g,z}S|^2-\Re(S)\rt)^N},\ \ \ \forall\ N\in\Z_+  
\ee  
where $C$ is a constant. Given that there is a finite distance in $\R^{n}$ between the non-Regge-like spin configuration $j_f$ and a Regge-like spin configuration, the above estimate implies the partial amplitude only gives a small contribution bounded by $o(\l^{-N})$ as $\l$ becomes large. However when the non-Regge-like spin configuration is close to a Regge-like one (in $\R^{n}$ distence), the above estimate may not imply $A_{\l j_f}(\ck)$ to be small, because $\min\lt(|\delta_{g,\xi,z}S|^2-\Re(S)\rt)$ may close to zero as the $j_f$ is close to be Regge-like, so that $\min\lt(|\delta_{g,\xi,z}S|^2-\Re(S)\rt)^{-N}$ becomes large and may cancel the decaying factor $\lt({1}/{\l}\rt)^N$. For example, since the spin-gap $\Delta j_f=\frac{1}{2\l}$, as a reasonable guess, $\min\lt(|\delta_{g,z}S|^2-\Re(S)\rt)$ for a non-Regge-like neighbor of a Regge-like spin may be of the order $1/\l$, which cancels the decaying behavior given by $\lt({1}/{\l}\rt)^N$. Therefore the generalized stationary phase analysis doesn't clarify how much a partial amplitude $A_{\l j_f}(\ck)$ with a generic non-Regge-like spin configuration contributes the spinfoam state-sum. But in order to analyze the spinfoam state-sum, we have to understand the contributions from all the partial amplitudes with both Regge-like and non-Regge-like spins, at least inside a neighborhood of $\mathring{j}_f$.

The above difficulty is from the fact that the non-Regge-like spins doesn't result in any critical point, which prevents us to write a closed formula to estimate the integral Eq.\Ref{Aj1} for all possible spins. The mathematical reason of the difficulty may be described as the follows: The integral Eq.\Ref{Aj1} are defined on an $n$-dim real space but with a complex action. The equations of motion are obtained by the variation of the complex action with $n$ real variables $x$, which results in $n$ complex equations, i.e. $2n$ real equations, for $n$ real variables. Then although the Heissian matrix $\delta^2_{x,x} S$ is assumed to be nondegenerate, the implicit function theorem doesn't implies there is always a solution for the $2n$ equations of motion, since the system is over-constrained. Furthermore, there are even some additional constraints coming from $\Re(S)=0$. 

However the problem over-constrained equations may be overcomed if one can complexify the space for integral, replacing the real variable $x$ to its complexification $z$ and analytic-continuing the action $S$ to the complexified space. Then the equations of motion $\delta_z S$ give $n$ complex equations with $n$ complex variables. The nondegenerate Heissian $\delta_{z,z}^2 S$ at a given critical point always implies a solution in the neighborhood by the complex version of implicit function theorem, at least when $S$ is an analytic function.  

Such an idea indeed works and well-studied in the mathematical literatures e.g. \cite{almostanalytic}, see also \cite{stationaryphase}. The machinery can even apply to the case that $S$ is $C^\infty$ functions, which admits (nonunique) almost analytic extension\footnote{An almost analytic extension $\tilde{f}$ of $f\in C^\infty(\R)$ in a neighborhood $\o$ satisfies (1) $\tilde{f}=f$ in $\o\cap\R$, (2) $|\partial_{\bar{z}}\tilde{f}|\leq C_N|\Im(z)|^N$ for all $N\in \Z_+$, i.e. $\partial_{\bar{z}}\tilde{f}$ vanishes to infinite order on the real axis. }. The result can be summarized in the following (Theorem 2.3 in \cite{almostanalytic}, see also Theorem 7.7.12 in \cite{stationaryphase})

\begin{Theorem} \label{almostanalytic}

Let $S(j,x)$, $j\in\R^k,\ x\in\R^N$, be an smooth function in a neighborhood of $(\mathring{j},\mathring{x})$. We suppose that $\Re\lt[S(j,x)\rt]\leq0$, $\Re \lt[S(\mathring{j},\mathring{x})\rt]=0$, $\delta_x S(\mathring{j},\mathring{x})=0$, and $\delta^2_{x,x}S(\mathring{j},\mathring{x})$ is nondegenerate. We denote by $\cs(j,z)$, $j\in\C^k,\ z=x+iy\in\C^n$ an (nonunique) almost-analytic extension of $S(j,x)$ to a complex neighborhood of $(\mathring{j},\mathring{x})$. The equations of motion $\delta_z \cs=0$ define an almost-analytic manifold $M$ in a neighborhood of $(\mathring{j},\mathring{x})$, which is of the form $z=Z(j)$. On $M$ and inside the neighborhood, there is a positive constant $C$ such that for $j\in\R^k$
\be
-\Re[\cs(j,z)]\geq C|\Im(z)|^2,\ \ \ \ z=Z(j) \label{recs}
\ee
We have the following asymptotic expansion for the integral
\be
\int e^{\l S(j,x)}\ u(x)\ \rmd x\sim e^{\l \cs\lt[j,Z(j)\rt]}\lt(\frac{1}{\l}\rt)^{\frac{N}{2}}\sqrt{\det\lt(\frac{2\pi i}{ \cs''\lt[j,Z(j)\rt]}\rt)}\sum_{s=0}^\infty\lt(\frac{1}{\l}\rt)^{s}\lt[L_{s}\tilde{u}\rt]\Big(Z(j)\Big)\label{expcs}
\ee
where $u(x)\in C^\infty_0(K)$ is a compact support function on $K$ inside the domain of integration. $N$ is the number of independent of $x$-variables, the same as the number of holomorphic $z$-variables. The differential operator $L_{s}$ is defined in the same way as in Theorem \ref{asymptotics} but operates on an almost analytic extension $\tilde{u}(z)$ of $u(x)$ and evaluating the result at $z=Z(j)$. The branch of the square-root is defined by requiring $\sqrt{\det\lt({2\pi i}\big/{ \cs''\lt[j,Z(j)\rt]}\rt)}$ to deform continuously to $1$ under the homotopy:
\be
(1-s)\frac{2\pi i}{ \cs''\lt[j,Z(j)\rt]}+s\mathbf{I}\in\mathrm{GL}(n,\C),\ \ \ \ s\in[0,1].
\ee
Note that the asymptotic expansions from two different almost-analytic extensions of the pair $S(j,x),u(x)$ are different only by an contribution bounded by $C_K\l^{-K}$ for all $K\in\Z_+$.

\end{Theorem}

\section{Analytic Extension of Spinfoam Action}\label{EFF}

When we apply the almost-analytic machinery to the spinfoam action Eq.\Ref{action}, we find the $S[j_f,g_{ve},z_{vf}]$ is actually an analytic function on a neighborhood of $\background$ (without touching a singularity), thus admits a unique \emph{analytic} extension. Indeed, as it was mentioned previously, the arguments of the logarithmics are the ratios of polynormials in the variables $\fz_{ve},\fa_{ve},\fb_{ve},\fc_{ve},\fd_{ve}$ and their complex conjugates, thus are ratios of the polynomials of their real and imaginary parts. Thus we can analytic extend the spinfoam action $S$ from the space of spinfoam configurations $\cm_{j,g,z}$ to its complexification $\cm_{j,g,z}^\C$ by replacing all the real variables (the real and imaginary part of spinfoam variables) by complex variables, or equivalently treating $\fz_{ve},\fa_{ve},\fb_{ve},\fc_{ve},\fd_{ve}$ and their complex conjugates as independent complex variables. The resulting spinfoam action $\cs$ is an analytic function in a complex neighborhood of $\background$ in $\cm_{j,g,z}^\C$.

We first make a precise definition of the complexified space $\cm_{j,g,z}^\C$. We define two types of group variables $(g_{ve},\tilde{g}_{ve})\in \Slc\times\Slc\simeq \mathrm{SO}(4,\C)$ parametrized by
\be
g_{ve}^t=\left(\begin{array}{cc}\fa_{ve} & \fb_{ve}\\ \fc_{ve} & \fd_{ve} \end{array}\right), \ \ \ 
\tilde{g}_{ve}^t=\left(\begin{array}{cc}\tilde{\fa}_{ve} & \tilde{\fb}_{ve}\\ \tilde{\fc}_{ve} & \tilde{\fd}_{ve} \end{array}\right)
\ee
as the complexifications of the $\Slc$ group variables. They give the complexifed Lorentz transformations on complexified Minkowski space. The spinorial representation of the complexified $\Slc$ consists of both dotted and undotted spinors, as the complexification of the spinorial representation of $\Slc$. Therefore we define two types of $\mathbb{CP}^1$ spinor $(z_{vf}^\a,\tilde{z}_{vf}^{\dot{\a}})\in \mathbb{CP}^1\times \mathbb{CP}^1 $ parametrized by
\be
z_{ve}^\a=\left(\begin{array}{c}1\\ \fz_{vf} \end{array}\right),\ \ \ \ \tilde{z}_{ve}^{\dot{\a}}=\left(\begin{array}{c}1\\ \tilde{\fz}_{vf} \end{array}\right)
\ee
The complexified space of spinfoam configurations $\cm_{j_f,g,z}^\C$ is a collection of the data $(j_f,g_{ve},\tilde{g}_{ve},z_{vf},\tilde{z}_{vf})$, where $j_f$ may be also taken as a complex variable.

The analytic extended action is given by 
\be
\cs=\sum_{(e,f)}j_f\lt[\ln\frac{\lt(g_{ve}^tz_{vf}\cdot\tilde{g}_{v'e}^t\tilde{z}_{v'f}\rt)^{2}}{ \lt( g_{v'e}^tz_{v'f}\cdot\tilde{g}_{v'e}^t\tilde{z}_{v'f}\rt) \lt(g_{ve}^tz_{vf}\cdot\tilde{g}_{ve}^t\tilde{z}_{vf}\rt)}+i\g \ln \frac{\lt( g_{ve}^tz_{vf}\cdot\tilde{g}_{ve}^t\tilde{z}_{vf}\rt)}{\lt( g_{v'e}^tz_{v'f}\cdot\tilde{g}_{v'e}^t\tilde{z}_{v'f}\rt)}\rt].\label{acaction}
\ee
where $Z_1\cdot\tilde{Z}_2\equiv \delta_{\a\dot{\a}}Z_1^\a \tilde{Z}_2^{\dot{\a}}$ which appears to be not manifestly Lorentz invariant because of the time-gauge imposed in the spinfoam model\footnote{See \cite{Lor} for a discussion about the Lorentz covariance of the EPRL spinfoam model.}. The analytic extended spinfoam action $\cs[j_f,g_{ve},\tilde{g}_{ve},z_{vf},\tilde{z}_{vf}]$ is a holomorphic function on the complexified space $\cm_{j,g,z}^\C$. When we impose the reality condition that 
\be
\tilde{g}_{ve}=\bar{g}_{ve},\ \ \tilde{z}_{vf}=\bar{z}_{vf}.
\ee
The analytic extended action reduce the usual spinfoam action Eq.\Ref{action} by identifying (the left-hand side is the variable in Eq.\Ref{acaction} and the right-hand side is the variable in Eq.\Ref{action})
\be
{g}_{ve}\sim {g}_{ve},\ \ \tilde{z}_{vf}\sim {z}_{vf}.
\ee

The continuous gauge freedom of the analytic extended action is described in the following: First of all at each vertex $v$ there is a $\mathrm{SO}(4,\C)$ gauge freedom. Given $(x_v,\tilde{x}_v)\in \Slc\times\Slc$, the action Eq.\Ref{acaction} is invariant under the following transformation
\be
g_{ve}\mapsto x_v^{-1} g_{ve},\ \ \ z_{vf}\mapsto x^t_vz_{vf}\nonumber\\
\tilde{g}_{ve}\mapsto \tilde{x}_v^{-1} \tilde{g}_{ve},\ \ \ \tilde{z}_{vf}\mapsto \tilde{x}^t_v\tilde{z}_{vf}
\ee
since the transformation leaves the combinations $g_{ve}^tz_{vf},\tilde{g}_{ve}^t\tilde{z}_{vf}$ invariant. Secondly, on each edge $e$ there is a $\Slc$ gauge freedom. Given $h_e\in\Slc$ we define $\tilde{h}_e=(h_e^{-1})^t$ and parametrize $h_e,\tilde{h}_e$ by
\be
h_e=\left(\begin{array}{cc}\mu_{e} & -\tilde{\nu}_{e}\\ \nu_{e} & \tilde{\mu}_{e} \end{array}\right),\ \ \tilde{h}_e=\left(\begin{array}{cc}\tilde{\mu}_{e} & -{\nu}_{e}\\ \tilde{\nu}_{e} & {\mu}_{e} \end{array}\right)
\ee
with $\mu_e\tilde{\mu}_e+\nu_e\tilde{\nu}_e=1$. Here we emphasize that $\mu_e,\tilde{\mu}_e$ are 2 independent complex variables and the same for $\nu_e,\tilde{\nu}_e$, so that both $h_e,\tilde{h}_e\in\Slc$. The analytic extended action $\cs$ is invariant under the following gauge transformation on the edge:
\be
g_{ve}^t\mapsto h_e g_{ve}^t,&&\tilde{g}_{ve}^t\mapsto \tilde{h}_e \tilde{g}_{ve}^t
\ee 
Inside each inner product, $h_e,\tilde{h}_e$ cancel each other by $h^t_e\tilde{h}_e=1$. Thirdly the following $\C\times\C$ complex scaling for each pair $(z_{vf},\tilde{z}_{vf})$
\be
z_{vf}\mapsto \l z_{vf},\ \ \ \tilde{z}_{vf}\mapsto \tilde{\l} \tilde{z}_{vf}, \ \ \forall\ (\l,\tilde{\l})\in\C\setminus\{0\}\times\C\setminus\{0\}.
\ee 

The equations of motion $\delta_{g,\tilde{g},z,\tilde{z}}\cs=0$ can be computed straight-forwardly. Unfortunately the resulting equations are complicated in the sense that it is not clear to us so far about the geometrical/physical interpretation of these equations. However it is clear to us that the equations of motion are a set of algebraic equations, which can be solved in principle at least in a neighborhood $\O^\C\in\cm^\C_{j,g,z}$ of $\background$ modulo gauge freedom, provided that the Hessian matrix is nondegenerate. The resulting solution defines an analytic manifold $Z(j)$ in $\O^\C$, where $Z(j)$ stands for $\lt(g_{ve}(j),\tilde{g}_{ve}(j),z_{vf}(j),\tilde{z}_{vf}(j)\rt)$ modulo gauge transformations. The background data is certainly a point on the manifold $Z(j)$ with $\mathring{Z}\equiv Z(\mathring{j})=\lt(\mathring{g}_{ve}(j),\mathring{\bar{g}}_{ve}(j),\mathring{\bar{z}}_{vf}(j),\mathring{z}_{vf}(j)\rt)$.

Formally we insert the solution $Z(j)$ into the asymptotic expansion Eq.\Ref{expcs}. We see that $\l \cs[j,Z(j)]\equiv \l\sum_fj_f\tilde{\cf}_f[Z(j)]$ is the leading order contribution of the spin effective action $\cw_\ck[\l j_f]$. The spin effective action can be written in the following way:
\be
\cw_\ck[\l j_f]=\l\cs[j,Z(j)]-\frac{N_{g,z}}{2}\ln\l+\half\ln{\det\lt(\frac{2\pi i}{ \cs''\lt[j,Z(j)\rt]}\rt)}+\ln\tilde{\mu}\lt[Z(j)\rt]+o(\l^{-1})\label{analycw}
\ee  
although we don't derive an explicit expression of $\cs[j,Z(j)]$ as a function of spins $j_f$. We do know that the $\cs[j,Z(j)]$ is an analytic function of $j_f$ in a neighborhood of $\mathring{j}_f$, since $\cs[j_f,g_{ve},\tilde{g}_{ve},z_{vf},\tilde{z}_{vf}]$ is a holomorphic function and $Z(j)$ is analytic in $\O^\C$. By an analytic power-series expansion of $\cs[j,Z(j)]$ we obtain that
\be
\cs[j,Z(j)]=\cs[\mathring{j},Z(\mathring{j})]+\sum_f\fs_f\frac{\partial\cs[{j},Z]}{\partial j_f}\Big|_{\mathring{j},\mathring{Z}}+\sum_f\fs_f\frac{\partial\cs[{j},Z]}{\partial Z}\Big|_{\mathring{j},\mathring{Z}}\frac{\partial Z(j)}{\partial j_f}\Big|_{\mathring{j}}+o(\fs_f^2)
\ee
where $\fs_f$ is the fluctuation $\fs_f=j_f-\mathring{j}_f$. From the asymptotic analysis of the spinfoam partial amplitude with Regge-like spins $\mathring{j}_f$, we know the expression of the following coefficients:
\be
\cs[\mathring{j},Z(\mathring{j})]=S[\mathring{j}_f,\mathring{g}_{ve},\mathring{z}_{vf}]=i\sum_f\g \mathring{j}_f\mathring{\Theta}_f,\ \ \ 
\frac{\partial\cs[{j},Z]}{\partial j_f}\Big|_{\mathring{j},\mathring{Z}}=i\g\mathring{\Theta}_f,\ \ \ 
\frac{\partial\cs[{j},Z]}{\partial Z}\Big|_{\mathring{j},\mathring{Z}}=0
\ee
which reproduces the result of coefficients $\cw_0,\cw_1^f$ in Eq.\Ref{SEA} at their leading orders (see Eqs.\Ref{cw0} and \Ref{cw1}). The $\ln\l$-term and $o(1)$-term of $\cw_0$ in Eq.\Ref{cw0} is also easily recovered in Eq.\Ref{analycw}, since $\cs''\lt[\mathring{j},Z(\mathring{j})\rt]=S''(\mathring{x})$ and $\tilde{\mu}\lt[Z(\mathring{j})\rt]=\mu(\mathring{x})$ by the (almost)-analytic extension. More importantly, by comparing the two expressions of $\cw_\ck[\l j_f]$ in Eqs.\Ref{SEA} and \Ref{analycw}, we find that the leading contributions of all the coefficients $\cw_n^{f_1\cdots f_n}$ in Eq.\Ref{SEA} are of $o(1)$. As a result all the coefficients $\cw_n^{f_1\cdots f_n}$ in Eq.\Ref{SEA} have the following asymptotic expansion:
\be
\cw_n=W_n^{(0)}+\lt(\frac{1}{\l}\rt)W_n^{(1)}+\lt(\frac{1}{\l}\rt)^2W_n^{(2)}+o(\l^{-3})
\ee
where for $n=0$ we write $W_0^{(0)}=i\sum_f\g \mathring{j}_f\mathring{\Theta}_f-\frac{N_{g,z}}{2\l}\ln{\l}$. It shows that our expectation in the discussion toward Eq.\Ref{cwn} is indeed correct. Moreover the analyticity of $\cs[j,Z(j)]$ implies that $\sum_{n=1}^\infty [W_n^{(0)}]^{f_1\cdots f_n}\fs_{f_1}\cdots\fs_{f_n}$ is convergent in a neighborhood of $\mathring{j}_f$.

There is another piece of information that we can extract from the almost-analytic machinery in Theorem \ref{almostanalytic}. We find that the action $\cs[j,Z(j)]$ is complex, whose real part is constantly non-positive in the neighborhood under consideration, i.e. there is a positive constant $C$ such that
\be
\Re\lt(\cs[j,Z(j)]\rt)\leq -C|\Im \lt(Z(j)\rt)|^2.\label{SZ}
\ee
It implies that there is an ``exponentially decaying'' factor in Eq.\Ref{expcs} except the Regge-like spin configurations:
\be
e^{\l\Re\lt(\cs[j,Z(j)]\rt)}\leq e^{-C\l|\Im \lt(Z(j)\rt)|^2}
\ee
If we make an power-series expansion of the analytic function $Z(j)$:
\be
Z(j)=Z(\mathring{j})+\sum_f\fs_f\partial_{j_f}Z(\mathring{j})+\half\sum_{f,f'}\fs_f\fs_{f'}\partial_{j_f}\partial_{j_{f'}}Z(\mathring{j})+o(\fs_f^3)\label{Zss}
\ee
where $Z(\mathring{j})$ is real since it recovers the background spinfoam variables certainly without complexification. Therefore we obtain the following expression of the decaying factor:
\be
e^{\l\Re\lt(\cs[j,Z(j)]\rt)}\leq e^{-C\l\lt|\Im \lt(\sum_f\fs_f\partial_{j_f}Z(\mathring{j})+\half\sum_{f,f'}\fs_f\fs_{f'}\partial_{j_f}\partial_{j_{f'}}Z(\mathring{j})+o(\fs_f^3)\rt)\rt|^2}
\ee
If the fluctuation $\fs_f=j_f-\mathring{j}_f\sim o(1)$, the factor $e^{\l\Re\lt(\cs[j,Z(j)]\rt)}$ is exponentially decaying, which results in that the corresponding partial amplitude $A_{\l j_f}(\ck)$ is bounded by $o(\l^{-N})$ for all $N\in\Z_+$. However if the fluctuation $\fs_f=j_f-\mathring{j}_f\sim o(\l^{-1/2})$, i.e. the non-Regge-like $j_f$ is close to a Regge-like spin configuration $\mathring{j}_f$, one can \emph{not} conclude that the factor $e^{\l\Re\lt(\cs[j,Z(j)]\rt)}$ decays as $\l$ large, because the $e^{-\l}$ behavior may be cancelled by $\fs_f\sim o(\l^{-1/2})$. It is in general not correct if one ignores the contribution in the state-sum from the partial amplitude with $\fs_f=j_f-\mathring{j}_f\sim o(\l^{-1/2})$ in the large spin regime.

\section{Implementation of Spin-Sum}

\subsection{$\l^{-1}$-Expansion and Flatness}

In this section we implement the sum over spins in the large spin regime. By the above asymptotic expansion of the spin effective action $\cw_\ck[\l j_f]$, we can write the spin-sum by\footnote{The spinfoam amplitude $A(\ck)$ under consideration here is actually $A_\tau$ in the previous context, since we only interested in the perturbative corrections at different critical points. The nonperturbative effects are taken into account by summing over critical points. $\tau$ is a cut-off function which makes $A(\ck)$ finite, and let us to zoom into the neighborhood of the background $\background$ to perform the perturbation expansion rigorously. It doesn't has any physical consequence to our analysis.}
\be
A(\ck)=\sum_{j_f=-\infty}^\infty d_{J_f}\t(j_f)\exp\cw_\ck[\l j_f]=\lt(\frac{1}{\l}\rt)^{\frac{N_{g,z}}{2}}\sum_{J_f\in\Z/2}d_{J_f} e^{\sum_f J_f\tilde{\cf}_f\lt[Z(J_f/\l)\rt]+o(1)}\t(J_f/\l)
\ee
where $0\leq\t(j_f)\leq 1$ is a smooth function of compact support located in $j_f\geq o(1)>0$. Since the summand is a compact support function on $\R^{N_f}$, we apply the Poisson resummation formula to the spin-sum:
\be
A(\ck)&=&\lt(\frac{1}{\l}\rt)^{\frac{N_{g,z}}{2}}\sum_{k_f\in\Z}\int_{\R^{N_f}}\lt[d_{J_f} \rmd J_f\rt] e^{\sum_f J_f\lt(\tilde{\cf}_f\lt[Z(J_f/\l)\rt]-4\pi ik_f\rt)+o(1)}\t(J_f/\l)\nonumber\\
&=&\lt(\frac{1}{\l}\rt)^{\frac{N_{g,z}}{2}-2N_f}\sum_{k_f\in\Z}\int_{\R^{N_f}}\lt[j_f\rmd j_f\rt]  e^{\sum_f \l j_f\lt(\tilde{\cf}_f\lt[Z(j_f)\rt]-4\pi ik_f\rt)+o(1)}\t(j_f)\label{universal}
\ee
For each branch $k_f$, we can study the integral use the stationary phase approximation, and we obtain the equation of motion:
\be
\tilde{\cf}_f\lt[Z(j_f)\rt]=4\pi i k_f,\ \  k_f\in\Z
\ee 
where we have used $\frac{\partial\cs\lt[j,Z(j)\rt]}{\partial Z}=0$ because of the equations of motion from analytic extended spinfoam action. On the other hand, By Eq.\Ref{SZ}, $\Re(\cs)=0$ implies $\Im(Z(j))=0$, i.e. $Z(j)$ reduces back to the critical data $\mathring{g}_{ve},\mathring{z}_{vf}$. Then the equation of motion reduces to
\be
\g\mathring{\Theta}_f=4\pi k_f,\ \  k_f\in\Z
\ee
Therefore all the critical points of $j_f$ are given by the critical configurations with $\g\mathring{\Theta}_f=0$ mod $4\pi\Z$, when we focus on the globally oriented and time-oriented critical configurations of Lorentzian geometry. This reproduces the flatness result given in \cite{flatness}. Such a flatness as the leading contribution in $\l^{-1}$ expansion is exactly resulting from the non-Regge-like fluctuation of $j_f$. Suppose we restrict the spin-sum only in the subset of Regge-like spin, as the leading contribution of the state-sum in $\l^{-1}$, we would obtain a sum of $e^{iS_{\mathrm{Regge}}}$ over the Regge-like spin configurations, instead of obtaining contributions only from flat (mod $4\pi\Z$) simplicial geometries. As a result the spinfoam state-sum in the large spin regime can be written as a sum over critical configurations $(j_f,g_{ve},z_{vf})$ with $\g\mathring{\Theta}_f=4\pi \Z$ (we set $\t(j_f)=1$ at these critical values of $j_f$):
\be
A(\ck)=\lt(\frac{1}{\l}\rt)^{\frac{N_{g,z}-3N_f}{2}}\sum_{(j_f,g_{ve},z_{vf})}j_f e^{I(j_f,g_{ve},z_{vf};\l)}
\ee
where the perturbative effective action $I(j_f,g_{ve},z_{vf};\l)$ is an asymptotic power-series in $\l^{-1}$, with the leading contribution of $o(1)$, i.e. the 1-loop $\ln\det(\cdots)$ contribution. $N_f$ denotes the number of triangles in $\ck$, and $N_{g,z}$ denotes the number of degrees of freedom in $g_{ve},z_{vf}$.

\subsection{Nontrivial Curvature from Re-expansion }

It is obvious that the flatness result $\g\mathring{\Theta}_f=0$ mod $4\pi \Z$ relies on the setting that $\l^{-1}$ is the only expansion parameter. However such a result may be modified by modifying the setting of expansion parameter. It is shown in the following that it is natural to have another expansion parameter $\g\mathring{\Theta}$ (background curvature) in addition to $\l$ (background spin). The 2-parameter expansion includes the non-flat curvatures into the effective degrees of freedom, while the expansion with respect to $\g\mathring{\Theta}$ is interpreted to be a low-energy expansion as being a curvature expansion. See Section 6 and \cite{han} for more physical motivations about the new expansion.

Let's write the spinfoam state-sum perturbatively at $\background$
\be
&&A_\t(\ck)=\prod_{f}d_{\mathring{J}_f}e^{\l\cw_0}\cz,\ \ \text{where}\ \ \cz=\sum_{\fs_f=-\infty}^\infty\prod_{f}\lt(1+\frac{2\l\fs_f}{d_{\mathring{J}}}\rt) e^{\l\sum_f\cw_1^f\fs_f+\l\sum_{f,f'}\cw_2^{f,f'}\fs_f\fs_{f'}+o(\fs^3)}\t\lt(\fs_f\rt),\label{As}\\
&&
\text{or}\ \ \ \cz=\sum_{\Fs_f=-\infty}^{\infty}\prod_{f}\lt(1+\frac{2\Fs_f}{2\l\mathring{j}_f+1}\rt) e^{\sum_f\cw_1^f\Fs_f+\frac{1}{\l}\sum_{f,f'}\cw_2^{f,f'}\Fs_f\Fs_{f'}+o(\l^{-2}\Fs^3)}\t\lt(\Fs_f/\l\rt)\label{wl}
\ee
where the leading contributions of all coefficients $\cw_n$ are of $o(1)$, and recall that $\t(\fs)$ is a smooth function of compact support from a partition of unity in the space of $j_f$. $\cw_0$ contains the Regge action of GR as the leading order contribution, $\cw_1$ contains the Regge deficit angle as the leading order contribution
\be
\cw_0=i\sum_f\g \mathring{j}_f\mathring{\Theta}_f-\frac{N_{g,z}}{2\l}\ln{\l}+o(\l^{-1}),\ \ \ \cw_1^f=i\g\mathring{\Theta}_f+o(\l^{-1}).
\ee
Then the spin-sum $\cz$ can be written in the following form by using the Poisson resummation formula:
\be
\cz
&=&2^{N_f}\sum_{k_f\in\Z}\int_{\R^{N_f}}\lt[\rmd \Fs_f\rt]\prod_{f}\lt(1+\frac{2\Fs_f}{2\l\mathring{j}_f+1}\rt) e^{i\sum_f\lt(\g\mathring{\Theta}_f-4\pi k_f\rt)\Fs_f+\frac{1}{\l}\sum_{f,f'}\cw_2^{f,f'}\Fs_f\Fs_{f'}+o(\l^{-2}\Fs^3)}\t\lt(\Fs_f/\l\rt)\nonumber\\
&=&(2\l)^{N_f}\sum_{k_f\in\Z}\int_{\R^{N_f}}\lt[\rmd \fs_f\rt]\prod_{f}\lt(1+\frac{2\l\fs_f}{2\l\mathring{j}_f+1}\rt) e^{i\l\sum_f\lt(\g\mathring{\Theta}_f-4\pi k_f\rt)\fs_f+\l\sum_{f,f'}\cw_2^{f,f'}\fs_f\fs_{f'}+o(\l\fs^3)}\t\lt(\fs_f\rt)\label{branch}
\ee
when we ignore the $o(\l^{-1})$-terms in $\cw_1$.

Now we assume the value of $\g\mathring{\Theta}_f-4\pi k_f\equiv\a \mathring{\Fx}_f$ is nonzero but very small for some certain $k_f$, where we make a new scaling parameter $\a\ll1$ and $\Fx_f\sim o(1)$. Then we have 
\be
\cz=(2\l)^{N_f}\sum_{k_f\in\Z}\int_{\R^{N_f}}\lt[\rmd \fs_f\rt]\prod_{f}\lt(1+\frac{2\l\fs_f}{2\l\mathring{j}_f+1}\rt) e^{i\b\sum_f\Fx_f\fs_f+\l\sum_{f,f'}\cw_2^{f,f'}\fs_f\fs_{f'}+o(\l\fs^3)}\t\lt(\fs_f\rt).
\ee
where we have defined $\b\equiv\l\a$ and treat $\b$ and $\l$ as two independent scaling parameter, as a reparametrization of the space of parameters $\a,\l$. $A_\t(\ck)$ is written as
\be
A_\t(\ck)=\prod_{f}d_{\mathring{J}_f}e^{i\b\sum_f\mathring{j}_f\mathring{\Fx}_f+o(1)}\cz.
\ee

We make the stationary phase approximation of the integrals in different branch $k_f$ with respect to the $\l$-scaling. The critical equations $\Re(\sum_{f,f'}\cw_2^{f,f'}\fs_f\fs_{f'}+o(\fs^3))=0$ and $\delta_{\fs_f}(\sum_{f,f'}\cw_2^{f,f'}\fs_f\fs_{f'}+o(\fs^3))=0$ are solved by $\fs_f=0$. Recall Theorem \ref{asymptotics}, the $\l^{-1}$ corrections are given by 
\be
\lt(\frac{1}{\l}\rt)^sL_s\lt[e^{i\b\sum_f\Fx_f\fs_f}\mu(\fs_f)\rt]_{\fs_f=0}\ \ \ \text{where}\ \ \ \mu(\fs_f)=\prod_{f}\lt(1+\frac{2\l\fs_f}{2\l\mathring{j}_f+1}\rt)\t\lt(\fs_f\rt)
\ee
Recall that $L_s$ is a differential operator of order $2s$, so we obtain the following power-counting
\be
\lt(\frac{1}{\l}\rt)^sL_s\lt[e^{i\b\sum_f\Fx_f\fs_f}\mu(\fs_f)\rt]_{\fs_f=0}=\sum_{r=0}^{2s}\frac{\b^r}{\l^s}f_{s,r}=\sum_{r=0}^{2s}{\a^r}{\l^{r-s}}f_{s,r}.
\ee
where we find that such an expansion is valid only when $\b^2\ll\l$, i.e.
\be
\a\ll\l^{-1/2}\label{alphall}
\ee
When such a perturbative expansion is valid, the critical configurations with nonzero $\g\mathring{\Theta}_f-4\pi k_f\equiv\a \mathring{\Fx}_f\ll o(\l^{-1/2})$ can contribute to the leading order approximation of $A(\ck)$:
\be
A(\ck)=\lt(\frac{1}{\l}\rt)^{\frac{N_{g,z}}{2}-2N_f}\sum_{(j_f,g_{ve},z_{vf})} j_f e^{i\b\sum_f{j}_f{\Fx}_f+\cdots}\label{A2}
\ee
where $(j_f,g_{ve},z_{vf})$ denote the critical configurations with $\g\mathring{\Theta}_f-4\pi k_f\equiv\a \mathring{\Fx}_f\ll o(\l^{-1/2})$, and ``$\cdots$'' stands for the contributions of $o(1)$ and $o(\b^r/\l^s)_{r\leq 2s}$. On the exponential $i\b\sum_f{j}_f{\Fx}_f$ is an analog of Regge action if we identify $\g j_f$ to be the area of the triangle $f$
\be
\exp \lt(i\b\sum_f{j}_f{\Fx}_f\rt)=\exp\lt( i\l\sum_f j_f\g\Theta_f\rt)
\ee
but here $\g\Theta_f$ is close to $2\pi\Z$ up to a difference much smaller than $o(\l^{-1/2})$.

\subsection{Semiclassical Low Energy Effective Action}

Let's try to understand the above expansion in a more detailed way. We focus on the integral with $k_f=0$ in $A(\ck)$. The integrals with nonzero $k_f$ can be analyzed in the same way. We apply the technique in Theorem \ref{almostanalytic} to the action 
\be
\l\cs\lt[\fs_f\rt]=i\l\sum_f\g\mathring{j}_f\mathring{\Theta}_f +\l\lt[\sum_fi\g\mathring{\Theta}_f\fs_f+\sum_{f,f'}\cw_2^{f,f'}\fs_f\fs_{f'}+o(\fs^3)\rt]
\ee
with $\Re(\cs)\leq 0$ by Eq.\Ref{recs}. 

If we consider the action $\cs[\fs_f]\equiv\cs[\g\mathring{\Theta}_f,\fs_f]$ where $\g\mathring{\Theta}_f$ is treated as a parameter here, we find 
\be
\g\mathring{\Theta}_f=0\ \ \text{and}\ \ \fs_f=0\ \ \Longrightarrow \ \ \Re(\cs)=0\ \ \text{and}\ \ \delta_{\fs_f}\cs=0
\ee
The critical point $\g\mathring{\Theta}_f=0,\fs_f=0$ for the action $\cs[\g\mathring{\Theta}_f,\fs_f]$ fullfils the assumption of the almost-analytic machinery in Theorem \ref{almostanalytic}. We apply the almost-analytic machinery to $\cs[\g\mathring{\Theta}_f,\fs_f]$ by analytic continuing $\fs_f$ to the complex variables. $\cs$ is an analytic function of $\fs_f$ then the equation of motion $\delta_{\fs_f}\cs=0$ gives an analytic manifold $\fs_f=Z_f(\g\mathring{\Theta}_f)$ at least locally. The integral in the branch $k_f=0$ in $\cz$ is expressed as an asymptotic expansion
\be
e^{i\l\sum_f\g\mathring{j}_f\mathring{\Theta}_f}\cz_{k=0}=e^{\l\cs\lt[\g\mathring{\Theta}_f,Z_f(\g\mathring{\Theta}_f)\rt]}\lt(\frac{1}{\l}\rt)^{\frac{N_f}{2}}\sqrt{\det\lt(\frac{2\pi i}{\delta^2_{\fs_f,\fs_{f'}}\cs\lt[\g\mathring{\Theta}_f,Z_f(\g\mathring{\Theta}_f)\rt]}\rt)}\lt[1+o\lt(\frac{1}{\l}\rt)\rt]\label{czk=0}
\ee
The leading effective action $\cs\lt[\g\mathring{\Theta}_f,Z_f(\g\mathring{\Theta}_f)\rt]$ has the property following from Eq.\Ref{recs} that
\be
\Re\lt(\cs\lt[\g\mathring{\Theta}_f,Z_f(\g\mathring{\Theta}_f)\rt]\rt)\leq -C\lt|\Im\lt(Z_f(\g\mathring{\Theta}_f)\rt)\rt|^2.\label{expect}
\ee
We can compute more concretely the expression of effective action $\cs\lt[\g\mathring{\Theta}_f,Z_f(\g\mathring{\Theta}_f)\rt]$ as a power series of $\g \mathring{\Theta}_f$. We expand the action $\cs[\g\mathring{\Theta}_f,\fs_f]$ at the 1st order solution (in $\g \mathring{\Theta}_{f}$) from equation of motion $\delta_{\fs_f}\cs=0$:
\be
\fs_f=-\frac{i}{2}\sum_{f'}(\cw_2^{-1})_{f,f'}\g \mathring{\Theta}_{f'}.
\ee 
If we define a short-hand notation: $y_f\equiv\fs_f+\frac{i}{2}\sum_{f'}(\cw_2^{-1})_{f,f'}\g \mathring{\Theta}_{f'}$, then the expansion of the action reads
\be
\l\cs\lt[\g\mathring{\Theta}_f,y_f\rt]&=&i\l\sum_f\g\mathring{j}_f\mathring{\Theta}_f+\l\lt[\frac{1}{4}\sum_{f,f'}(\cw_2^{-1})_{f,f'}\g \mathring{\Theta}_{f}\g \mathring{\Theta}_{f'}+o\lt((\g \mathring{\Theta}_{f})^3\rt)\rt]+\nonumber\\
&&\l\lt\{\sum_f\lt[o\lt((\g \mathring{\Theta}_{f})^2\rt)\rt]y_f+\sum_{f,f'}\lt[2\cw_2^{f,f'}+o\lt(\g \mathring{\Theta}_{f}\rt)\rt]y_fy_{f'}+ o(y^3_f)\rt\}
\ee 
The equation of motion $\delta_{y_f}\cs\lt[\g\mathring{\Theta}_f,y_f\rt]$ with respective to $y_f$ gives an $o((\g \mathring{\Theta}_{f})^2)$ correction to the original approximating solution $y_f=0$, i.e. 
\be
y_f=o((\g \mathring{\Theta}_{f})^2)\ \ \ \text{or}\ \ \ \fs_f=-\frac{i}{2}\sum_{f'}(\cw_2^{-1})_{f,f'}\g \mathring{\Theta}_{f'}+o((\g \mathring{\Theta}_{f})^2)
\ee
If we expand the action $\cs[\g\mathring{\Theta}_f,\fs_f]$ at the new approximating solution and iterating the above procedure, we approximate the exact solution of $\delta_{\fs_f}\cs[\g\mathring{\Theta}_f,\fs_f]=0$ better and better and obtain the exact solution $\fs_f$ as a power-series of $\g\mathring{\Theta}_f$
\be
\fs_f=Z_f(\g\mathring{\Theta}_f)=\sum_{n=1}^\infty \a_{f,f_1,\cdots,f_n}\g\mathring{\Theta}_{f_1}\cdots\g\mathring{\Theta}_{f_n}
\ee
where the series has a finite convergence radius since we know that $Z_f(\g\mathring{\Theta}_f)$ is analytic. 

Evaluating the action $\cs[\g\mathring{\Theta}_f,\fs_f]$ at this exact solution gives 
\be
\l\cs\lt[\g\mathring{\Theta}_f,Z_f(\g\mathring{\Theta}_f)\rt]&=&i\l\sum_f\g\mathring{j}_f\mathring{\Theta}_f+\l\lt[\sum_{n=2}^\infty \b^{(n)}_{f_1,\cdots,f_n}\g\mathring{\Theta}_{f_1}\cdots\g\mathring{\Theta}_{f_n}\rt]
\ee 
where the quadratic order coefficient is given by
\be
\b^{(2)}_{f_1,f_2}=\frac{1}{4}(\cw_2^{-1})_{f_1f_2}.
\ee
Therefore as $\g\mathring{\Theta}_f$ is small, 
\be
\cs\lt[\g\mathring{\Theta}_f,Z_f(\g\mathring{\Theta}_f)\rt]=i\sum_f\g\mathring{j}_f\mathring{\Theta}_f+\frac{1}{4}\sum_{f,f'}(\cw_2^{-1})_{f,f'}\g \mathring{\Theta}_{f}\g \mathring{\Theta}_{f'}+o\lt((\g \mathring{\Theta}_{f})^3\rt).
\ee

Following the same procedure for the $k_f\neq0$ branches in Eq.\Ref{branch}, we obtain in general
\be
&&\cs_k\lt[\g \mathring{\Theta}_{f}-4\pi k_f,Z_f(\g \mathring{\Theta}_{f}-4\pi k_f)\rt]\nonumber\\
&=&i\sum_f\g\mathring{j}_f\mathring{\Theta}_f+\frac{1}{4}\sum_{f,f'}(\cw_2^{-1})_{f,f'}\lt(\g \mathring{\Theta}_{f}-4\pi k_f\rt)\lt(\g \mathring{\Theta}_{f'}-4\pi k_f\rt)+o\lt((\g \mathring{\Theta}_{f}-4\pi k_f)^3\rt)
\ee 
In general $\cs_k$ has a negative real part coming from the terms of quadratic and higher order in $(\g \mathring{\Theta}_{f}-4\pi k_f)$. By Eq.\Ref{expect}, the exponential $e^{\l\cs_k}$ in the asymptotic expansion Eq.\Ref{czk=0} for generic $k$ decays exponentially unless $\g\mathring{\Theta}_f$ from the background data $\background$ is close to one of $\{4\pi k_f:k_f\in\Z\}$. The non-decaying $e^{\l\cs_k}$ requires the negative real part of $\l\cs_k$ doesn't scale to be large by $\l\gg1$, which gives the nontrivial restriction to $\g \mathring{\Theta}_{f}$, i.e. for a constant $C\sim o(1)$
\be
\lt|\g \mathring{\Theta}_{f}\rt|\leq C\l^{-1/2}\ \ \ \text{mod}\ \ 4\pi \Z. \label{bound}
\ee
which improves the bound Eq.\Ref{alphall}.

If we assume $\g\mathring{\Theta}_{f}$ is small and the bound Eq.\Ref{bound} is satisfied, the above analysis let us obtain an perturbative effective action for the spinfoam state-sum $A(\ck)$:
\be
A(\ck)=\prod_{f}d_{\mathring{J}_f}\lt(1+\mathring{j}_f^{-1}Z_f(\g\mathring{\Theta}_f)\rt)e^{\l I_{\mathrm{eff}}\background}\label{result}
\ee
where the leading order contribution to the effective action $I_{\mathrm{eff}}$ is given by a power-series of $\g \mathring{\Theta}_{f}$
\be
I_{\mathrm{eff}}\background=i\sum_{f}\g \mathring{j}_f \mathring{\Theta}_{f}+\frac{1}{4}\sum_{f,f'}(\cw_2^{-1})_{f,f'}\g \mathring{\Theta}_{f}\g \mathring{\Theta}_{f'}+o\lt((\g \mathring{\Theta}_{f})^3\rt)-\frac{N_{g,z}-N_f}{2\l}\ln\l+o\lt(\frac{1}{\l}\rt).
\ee
We have neglected the $k_f\neq0$ integrals in Eq.\ref{branch} since they are exponentially decaying. Recall Eq.\Ref{CW2} for the expression of $\cw_2^{f,f'}$. Although $\cw_2^{f,f'}$ is local in $f,f'$, $(\cw_2^{-1})_{f,f'}$ is nonlocal. 

$\mathring{\Theta}_{f}$ is the deficit angle of the background simplical geometry from $\background$. Thus the above expression of effective action is a curvature expansion whose leading order is the Regge action.

\section{Discussion}

The above analysis studies the asymptotic behavior of spinfoam state-sum model by taking into account the sum over spins in the large spin regime. We mainly focus on the contributions from the globally oriented and time-oriented critical configurations of Lorentzian geometry. The analysis shows that:

\begin{enumerate}

\item If we only consider a single scaling parameter $\l$, which is the large spin parameter, the leading contributions to the spinfoam state-sum $A(\ck)$ are only given by the critical configurations with their deficit angle satisfying
\be
\g\Theta_f=0\ \ \text{mod}\ \ 4\pi\Z.
\ee

\item However if analyze the situation more carefully with $\g\Theta_f\ll 1$ mod $4\pi\Z$, a different expansion of $A(\ck)$ should be employed by using two scaling parameters $\l$ and $\g\Theta_f$. Such a expansion is valid only when $\g\Theta_f\leq \l^{-1/2}$ mod $4\pi\Z$.

\item The new 2-parameter expansion gives a curvature expansion of the semiclassical low energy effective action from the spinfoam model. In the expansion of effective action, the leading contribution is the simplicial Einstein gravity (Regge action), while the UV modifications of Einstein gravity appear as subleading high-curvature corrections.

\end{enumerate}

\noindent
Let's focus on the branch $k_f=0$, a small $\g {\Theta}_{f}=\a\Fx_f \leq o(\l^{-1/2})$ may come from two possibilities in particular (the combinations of the two are also possible): 

\begin{itemize}

\item If we require a small Barbero-Immirzi parameter e.g. $\g\ll\l^{-1/2}$ then a finite $\mathring{\Theta}_{f}\sim o(1)$ is admitted, i.e. we thus identify $\a=\g$ and $\Theta_f=\Fx_f$. Then the leading contribution Eq.\Ref{A2} gives the quantum Regge calculus with a discrete functional integration measure on the space of critical configurations (equivalent to the discrete cotetrads). If we make a further stationary phase approximation with respect to $\b$-scaling, the leading order contributions are given by the discrete cotetrad satisfying the discrete Einstein equation in Regge calculus.  

The requirement of a small $\g$ has already been proposed in the context of spinfoam graviton propagator computation \cite{propagator,3pt}, see also \cite{claudio}. A detailed and systematic study of this possibility is presented in \cite{smallgamma}. 

\item We assume the Barbero-Immirzi parameter $\g\sim o(1)$, and consider a critical configuration $(j_f,g_{ve},z_{vf})$ with the deficit angle $\Theta_f\sim \a \Fx_f \leq o(\l^{-1/2})$ (the deficit angles are scaled to be small by $\a$). The simplicial geometry from the critical data approximates a continuum geometry with a typical curvature radius $\rho$. The (dimensionful) averaged edge-length is $L$\footnote{In the geometrical interpretation of the spinfoam critical data, we assume the area $\g j_f$ is measured in the unit $\l\ell^2_P$, which leads to (1) the area spectrum $A_f=\g J_f \ell_P^2$ coincides with the result from canonical LQG in the large spin regime; (2) the Regge action is obtained by $\l\sum_f\g j_f\Theta_f=\frac{1}{\ell_P^2}\sum_fA_f\Theta_f$.}. In the continuum limit, the edge-length is much larger than the curvature radius. Then the Regge deficit angle can be written approximatly by \cite{FFLR}
\be
{\Theta}_{f}=\frac{\text{Area}(f^*)\ell_P^2}{\rho^2}\lt[1+o\lt(\frac{L^2}{\rho^2}\rt)\rt] \sim o(\l)\frac{\ell_P^2}{\rho^2}
\ee
where $\text{Area}(f^*)$ is the area of the face dual to the triangle $f$. Therefore if we keep Barbero-Immirzi parameter $\g\sim o(1)$,  
\be
{\Theta}_{f}\leq o(\l^{-1/2})\ \ \Longleftrightarrow\ \ \rho^2\geq\l^{3/2}{\ell_P^2}\label{bound}
\ee
i.e. the typical curvature radius of the background geometry should be much larger than the Planck length. If the large background spin parameter $\l\sim 10^{4}$, then the requirement from Eq.\Ref{bound} for the curvature radius is $\rho\geq 10^3\ell_P\sim 10^{-32}m$. It means that a large range of nontrivial curvature is still admitted as the background geometry, even when $\g\sim o(1)$. The expansion parameters contains the scaling $\a$ of deficit angle, so the expansion may be understood as a curvature expansion. 

In such an approximation, only the critical configurations with small deficit angles contribute in the leading order. One needs a large triangulation in order to approximate a curved geometry. Such a situation perhaps relates to the regime where the model approximate GR on the continuum. 

\end{itemize}

The analysis here mainly focus on the critical configurations of globally oriented, time-oriented Lorentzian geometry. However for the other critical configurations (with $\eps=-1,\sgn(V_4)=1$), the analysis can be carried out essentially in the same way \cite{smallgamma}. Eq.\Ref{universal} is a universal formula valid independent of the choice of critical configurations. Here we list the critical value of ${\cf}_f$ at different type of critical configurations classified in \cite{hanPI,HZ}:

\begin{center}
\begin{tabular}{| l || c |}
\hline
  & ${\cf}_f$\ \ \ \\ \hline \hline
\ Lorentzian Time-Oriented&\ \ $-i\,\eps\, \sgn(V_4)\g\Theta_f$\  \\ \hline
Lorentzian Time-Unoriented&\ \ $-i\,\eps\lt[ \sgn(V_4)\g\Theta_f+\pi\rt]$\  \\ \hline
\ \ \ \ \ \ \ \ \ \ \ Euclidean &\ \ $-i\eps \big[\sgn(V_4)\Theta^E_f+\pi n_f\big]$  \\ \hline
\ \ \ \ \ \ \ \ \ \ \ \ \ Vector &\ \ $i\Phi_f$    \\ \hline
  \end{tabular}\\
  \vspace{0.4cm}
  \underline{Table 1}.
\end{center}
where $\Theta^E_f$ is the deficit angle in Euclidean geometry, $n_f=0,1$, and $\Phi_f$ is the vector geometry angle. The generalization of the previous results to other critical configurations can be done straight-forwardly by replacing the quantity $\g\Theta_f$ in the previous discussion by other critical values of ${\cf}_f$.


\section*{Acknowledgements}

The author would like to thank H. Haggard, S. Speziale, A. Riello, C. Rovelli, and M. Zhang for many helpful discussions. He also would like to thank Y. Ma for the invitation to visit the Center for Relativity and Gravitation, Beijing Normal University, where a part of this research work is carried out. The research leading to these results has received funding from the People Programme (Marie Curie Actions) of the European Union's Seventh Framework Programme (FP7/2007-2013) under REA grant agreement No. 298786.

\end{document}